# Beyond The Desktop Spreadsheet

Gordon Guthrie, Stephen McCrory
hypernumbers.com
gordon@hypernumbers.com, stephen@hypernumbers.com

**ABSTRACT**

Hypernumbers is a new commercial web-based spreadsheet. It addresses several risk factors in deploying spreadsheets.

Traditional risk management of spreadsheets has focused on *run-time* risk - incorrect formulae, accidental overwriting of formula, unapproved adaption of the spreadsheets structure by an unapproved person and so on. The sources and lifecycles of spreadsheet errors are by now quite well known

Current risk management has three major mechanisms to address risk. Firstly, to instrument spreadsheets using their scripting capabilities to impose certain behaviours on them. Secondly, to minimize risk by using operating procedures and standards which are shown to be less error-prone that ad-hoc spreadsheet development methods. These are training intensive and hard to police. Thirdly, to inspect and audit spreadsheets in operation.

Hypernumbers is attempting to address spreadsheet risk in two radically new ways. The core approach is not to mitigate risk but to engineer out risky activities.

The curtilage of the spreadsheet has been extended to include much of the operational environment in which conventional spreadsheets are used. This process absorbs and subsumes much of the 'best practice' and end-user disciplines that are conventionally used to address risk.

Traditional spreadsheets have no clear barrier between *data* and *programme instructions* – in the Turing sense – or between *input data* and *output information*. Hypernumbers enables those barriers to be simply and reliably re-imposed, thus draining off a whole category of run-time errors. This separation allows spreadsheet usage to be split into two distinct phases – development, where audit and testing can reduce errors, and deployment, where what would be risky practices in other spreadsheet paradigms are simply engineered out.

Using these techniques, 'applications' consisting of 10,000+ spreadsheets used by tens of users can be, and have been, safely built.

## 1  A BRIEF HISTORY OF SPREADSHEETS

Spreadsheets arose in the swansong days of the mainframe and really exploded only with the rise of personal computer [ROIZEN, 2010]. The run-time environment was un-networked personal computers in a single user mode.

The early spreadsheet war was decisively won by Microsoft Excel, and the second phase began – a battle between proprietary software and the open source spreadsheets. This was, from a technical perspective, a re-run of the previous war, as all the open source contenders (including what became Open Office) had their roots in the un-networked, single user world.

The third war is in progress: a battle between web-enabled desktop spreadsheets and their earth-bound counterparts – the collaboration wars. Collaboration being, in the context of online spreadsheets, narrowly defined as two or more people editing the same spreadsheet sequentially.

This common architectural history explains the commonality of solutions to spreadsheet risk across all the contenders.

The conventional spreadsheet is a networked object – shared inputs, shared outputs – which is distributed via a variety of ad-hoc mechanisms because it is based on non-networked technologies.

## 2  HYPERNUMBERS – AN OVERVIEW

Hypernumbers resembles a spreadsheet in the way that a stick insect resembles a stick – it is a purposeful, but misleading resemblance. A better example would be the telephone dialer on a smart phone. A smart phone is a computer pretending to be a mobile phone, and a mobile phone is a two-way radio pretending to be a telephone. However, across the technological changes of the last century there has been a consistent and long-lived user mental-model – the dialer, the core user interface into the world of telephony.

The intention of Hypernumbers has been to enable the end-user to do new things, whilst preserving and building on the familiar mental model of a spreadsheet.

Hypernumbers aims to address the common problems arising from the ad-hoc distribution of spreadsheets (file systems, e-mail, etc, etc) by building on a networked base that eliminates them as a class of error.

## 3  THE PROBLEMS

### 3.1  Overview

This section will enumerate a set of common spreadsheet problems to provide a sound analytical basis for the discussion of risk management approach developed so far in Hypernumbers.

### 3.2  Co-ordination

Spreadsheets are typically used for co-ordination of activities, often by aggregation. This is usually achieved by using shared folders, collaborative technologies like Google Docs or Sharepoint, or sometimes by e-mail.

The negotiation of locking of spreadsheets for update is usually manual and tiresome.

### 3.3 Proliferation

Proliferation takes two characteristic forms.

Individual spreadsheet files are duplicated, 'the one on the work server', 'the one I emailed home to work on at the weekend', 'the one I sent you for your input'.

Individual departments or work units use spreadsheets to build ad-hoc operational systems. These departmental spreadsheets ecosystems overlap in scope and become impossible to manage.

### 3.4 Permission Management

The conventional desktop spreadsheet has very primitive permission management – it tends to be all or nothing – you can edit the whole spreadsheet or not see it. There is a password protection layer built into modern spreadsheets but it is ad-hoc and not based role-based access. In addition, if you can have physical possession of a spreadsheet, but not the password, you can eventually force access to it. It is not true permission management.

### 3.5 Version Control

The problems of version control are close cousins of the problems of proliferation. Publication and management of templates is difficult to do properly. Individual spreadsheets have no sense of their own history (am I version 1.0 or version 1.1?). Versioning has to be imposed by external systems, or operational discipline, and both are error prone.

### 3.6 Verification

Spreadsheets contain an admixture of *input data*, *programme instructions* (formulae) and *output data*. This makes it hard for verification to be sustained. The person who develops a spreadsheet may be appropriate trained, the template so produced can go through an audit process, but ultimately the end-user can then trample on that good work. The lack of access control and audit means that anytime a spreadsheet has 'been passed around' its correctness is automatically suspect. This makes verification somewhat like painting the Forth Railway Bridge, the end is only the beginning again.

### 3.7 Audit And Data Management

Audit is hard to do with traditional spreadsheets. As users are not identified intrinsically in the interaction with the spreadsheet; it is well-nigh impossible to build a picture of *what this person did to that spreadsheet* let alone *what this person did to those spreadsheets*.

Because the spreadsheet file is an uncontrolled object it is hard to ensure that it is backed up, reliably deleted and other conventional data management operations.

# 4  ADDRESSING THE PROBLEMS

## 4.1 Introduction

This paper contains just a short overview of the features of the Hypernumbers platform that pertain to addressing common problems in spreadsheet deployment. A fuller account can be found elsewhere [McCRORY, GUTHRIE, 2011].

## 4.2  Business Architecture (And Technical Architecture)

### Overview

The basic framework adopted to address these problems was a complete platform business architecture based on a new technical architecture.

### Web Centralisation

The core technical architecture chosen was a web-based platform with role-based access.

Spreadsheet proliferation is eliminated by making the spreadsheets accessible from anywhere. This aspect is not substantially different from the Google Docs or Sharepoint approach to centralisation. Central role-based access, however, makes it possible to build a set of sheets where Alice can edit her sales data, Bob can edit his marketing data, and Charlie can edit his fulfilment data, the whole aggregating seamless, and in an error free manner, to a report for the CEO.

### Hierarchical Structure

Traditional spreadsheets, both desktop and web, are organised as a set of tabs (siblings). By contrast it was decided that Hypernumbers sheets should be organised in a hierarchical space that resembles a file system. A particular spreadsheet might have the URL:
`http://example.com/accounts/2011/invoices/inv00000001/`

This can be thought of as the equivalent of a spreadsheet called `inv0000001.xls` in directory `accounts -> 2011 -> invoices`

The advantage of this organisational structure is that it enables a whole class of special queries. It makes it possible to write functions that correspond to *add up all the unpaid invoices* or *show me a list of all invoices that are more than 30 days overdue*.

A traditional spreadsheet can be considered in terms of x-y coordinates, 5 columns along 6 rows down. These novel queries, z-queries, operate in 'the third dimension', the z-dimension – they can be thought of as a *go through this pile of spreadsheets on my desk and add the totals up*.

### An Operational Platform

To address many of the basic management and housekeeping risks it was decided to deliver Hypernumbers as a PaaS – <u>P</u>latform <u>A</u>s <u>A</u> <u>S</u>ervice. The platform was defined to include certain intrinsic capabilities (daily backup, clustered high-availability service, central log-on and authentication, automatic logging) that would drain some of the swamps of operational risk associated with large spreadsheet bases.

**Problems Addressed**

The business and technical architecture of Hypernumbers was intended to addresses the following problems in whole or in part:
- proliferation
- audit and data management

### 4.3 Views And Permissions

**Overview**

There are three core roles in spreadsheet use:
- the maker – the person who designs the business logic and structure and creates the formulae
- the data inputer – the person (or system) who provides or inputs data to drive the business logic
- the output consumer – the person (or system) who receives and acts on the outputs of the spreadsheet

A decision was taken to bake these three roles into the structure of Hypernumbers. These three roles thus have a dedicated *view* of the data, respectively:
- the spreadsheet view
- the wikipage view
- the webpage view

In addition there are a couple of minor views:
- the table view
- the log view

**Spreadsheet View**

The maker can use a spreadsheet view of a particular page – it is familiar to all spreadsheet users with the appropriate cell and selection behaviours, keyboard short cuts, wizards and so on and for forth.

**Webpage View**

The output consumer can use a webpage view of a particular page – nothing can be changed, no spreadsheet operations can be performed. As this was regarded as a fundamental activity, it was decided to make creating the webpage views as easy as possible – the process was finally reduced to a single click.

**Wikipage View**

Data imputers can be provided with a wiki page. The maker can mark cells on a page as wiki page inputs (either simple text boxes or drop down menus). When the page is rendered as a wikipage these cells (and these cells only) can be edited by the end user. A decision was taken to make it easy to turn spreadsheets into web forms with a single click.

To keep the roles separate, the wikipage view also entirely prevents the user from making any structural changes (row and column insertions or deletions).

**Tables**

A common use case of spreadsheets is as primitive tabular data – each row is considered 'a thing'. The table view is simple row editor designed to expedite this usage.

Because this is a pervasive use case, a simple forms construct was developed which makes their creation easy.

**Logs**

It was decided that forensic activities should be automatic and not optional or bolted on.

As a result, by default all user actions are logged – in two different respects. There is a log view which provides real time access to the historical values of particular cells on particular pages which makes it possible to reconstruct the history of a particular value.

There is another narrative log which documents the interactions of a particular user across multiple pages. This is not yet integrated into the body of the system, but provides detailed forensic audit information.

**Permissions**

From a technical perspective permissions *could* be granular down to *particular actions* on *particular cells*. However usability experiments found that this was impractical. Simple spreadsheet pages would expose upward of 15,000 securable API points *in the viewable portion* of the page.

As a result of these tests a decision was taken that permissions should be applied at a view level. If a person can see a particular view of a particular page then they can perform all the actions that that view implies.

**Problems Addressed**

The intention was that the views and permissions functionality of Hypernumbers should address the following problems in whole or in part:
- co-ordination
- permission management
- audit

Views perform the role of separating out the various classes of user. Permissions ensure that the various classes are locked down. This combination is intended to address directly the bulk of common spreadsheet risks.

### 4.4 Functions

**Overview**

At the core of the spreadsheet model is the common programming precept of putting `=function(arg1, arg2,...)` *in a cell*.

Hypernumbers have taken as a design principle Blaise Pascal's famous dictum [PASCAL, 1656-1657]: *je n'ai fait celle-ci plus longue parceque je n'ai pas eu le loisir de la faire plus courte*[1].

Hypernumbers endeavoured to implement this precept in all respects. Considerable effort has gone into ensuring that all the various types of user needs, from web controls to database queries, html menus to debugging, all can be fulfilled by invoking functional expressions in cells.

**Layout**

Layout items for webpages and wikipages are implemented in functions. For example:
`=html.box.4x8("Title", "This is the body text", "Footer")`

This creates a styled box 4 standard cells wide and 8 standard cells high with a title, a body and a footer. The set of layout functions will be expanded to include standard web idioms like tabbed boxes, accordion boxes, slideshows and the like.

**Navigation**

Navigation on web pages is provided by menus. These are created by functions with the following syntax:
`=html.menu(`*ref to sub-menu1*`, `*ref to sub-menu2*`, ....)`

Other common web navigational items like crumb trails are provided by other functions. Future work will include making menu functions that read the underlying page and permissions structure of the Hypernumbers website to create dynamic menus.

**Business Logic**

At the core of the business logic is a large set of 100+ Excel-compatible functions. The current set of implemented functions is based on the ODF Small Group [OASIS OPEN DOCUMENTS, 2011], with some additional functions from the Medium and Large Group.

These have the same parameters (and throw the same errors) as their Excel and Open Office equivalents.

**Forms And Transactions**

Various elements for traditional web forms can be specified by functions in cells, for instance, radio boxes, text input boxes, dropdown lists and so on. These can be grouped into transactions by parameters. This approach enables data input to be managed, aggregated and validated without tedious audit. Because the functions are themselves programmable the transactions exposed to the end user are themselves programmable, providing a flexible, yet secure mechanism for handling data input.

---

[1] I have made this letter longer than usual, because I lack the time to make it short

**Database**

Database-like functionality is provided by so-called z-functions. Some of these take the form of additional parameter types for conventional functions. Consider the function
`=sum(..., ...)`

It behaves as you would expect:
`=sum(1, 2, a3, !page!a4)`

The reference to other sheets can also be by a standard 'web-style' addressing scheme where the Lotus-1-2-3 bangs are replaced by forward slashes:
`=sum(1, 2, a3, /page/a4)`

The page reference approach is totally generalised to the arbitrarily deep hierarchy of hypernumbers using a conventional web syntax with `../` being the parent page and `./` being the current page.

In addition to all this (mostly) familiar syntax, sum can also take z-references:
`=sum(1, 2, a3, /some/page/[a1 > 44]/b7)`

The expression `/some/page/[a1 > 44]/b7` is a z-expression. Imagine there are 4 pages on a Hypernumbers site:
`/some/page/bleh/`
`/some/page/blah/`
`/some/thing/blurg/`
`/another/page/`

The z-expression would be evaluated segment by segment against these pages to see if they match:
- the first three match the segment `/some/`
- the first two then go on to match the second segment `/page/`
- at this point the strange segment `/[a1 > 44]/` is applied. The brackets `[...]` simply indicate that it is a z-segment. The expression `a1 > 44` is evaluated on each page. If it returns `true` that page matches, if it returns `false` it doesn't. The values of the cell `b7` on every matching page is then included in the `sum`

The expressions in z-queries are simply valid excel-compatible spreadsheet expressions that evaluate to either `true` or `false`. They can use the full panoply of functions (with some minor exceptions). So for instance expressions like `[if(or(a1 > 0, b7 = "failed"), true, false)]` are valid z-segments.

These database functions (when combined with structural functions and templates) enable complex applications to be built.

**Integration**

Because web spreadsheets should be first-class web citizens there are functions like:
`=facebook.comments(`*`facebook_app_id`*`)`

This puts integrated facebook comments on webpages or wikipages. Other integration functions for twitter, and such-like web properties are available.

**Structural Functions**

There are a number of structural functions. A typical one would be:
`=create.button(`*`buttontitle`*`, `*`params`*`)`

This function takes a specification of a webpage to create. It typically builds it from a template, under a structural name and then forwards the user to that page.

An example of a specification is:
`=create.button("Prepare New Day", "/some/page/[blank, date, yyyy]/[blank, date, mm]/[day_sheet, date, dddd]/")`

This would create the following pages:
`/some/page/2011/`             from a `blank` template
`/some/page/2011/apr/`         from a `blank` template
`/some/page/2011/apr/21/`      from the `day_sheet` template

The person who clicked the button would be redirected to the page:
`/some/page/2011/apr/21/`

More sophisticated parameters enable the manipulation of views and groups at run time, as well as the creation of more complex sets of subpages.

**Problems Addressed**

These functions play their part in making it possible to address the following problems:
- co-ordination
- proliferation

None of them address these problems directly, but between them they make it possible to build sophisticated web-based applications that are usable by multiple people simultaneously.

However the primary purpose of the functional set is not to directly solve problems, but to make it as easy as possible for existing spreadsheet users to extend their repertoire of actions by presenting what are logically new activities in an already understood and mastered form.

**4.5 Templates And Structural Change**

**Overview**

In order to address issues pertaining to checking and verification of spreadsheets a decision was taken to make templates the basic units of development. They work the same way as templates in desktop spreadsheets – except their invocation is trivially programmable.

**Template Creation**

Templates are created by simply saving spreadsheet pages as templates. As well as saving the data, formulae and formatting the template includes special attributes (whether a cell is a wiki-editable cell for instance) as well as permissions and views. A page can be created as a wiki page for the sales department. All pages created from that template will then by default be wiki pages for the sales department.

**Structural Functions**

There is a need to enable repeatable abstractions in the creation of spreadsheets [PEYTON-JONES, BLACKWELL, BURNETT, 2003]. Templates can be used to fulfil *part* of this need if they can be invoked in a programmatic manner.

To this end, structural functions can be used to create buttons on webpages or wikipages which then instantiate new spreadsheets from templates at named places in the hierarchy. z-queries can then be used to aggregate the information captured in those templates, or to populate navigational items like html menus.

**The Development Process**

This use of templates and structural functions was designed to enable a novel development process to be implemented. An application can be built in spreadsheets. As part of the development process it can be audited and tested extensively and subject to quality measures.

By contrast it can be deployed as *an application*. In normal daily use the end-users only see webpages or wikipages – they can neither see, nor change, the underlying 'programme' aspects. They can enter data into the application and see the appropriate outputs. They can create new instances of wikipages and webpages by pressing structural buttons. The results of their interaction with those new pages can be automatically aggregated and summarise up the hierarchy.

The problem of data integrity *at run time* then becomes one of validating data entry. Was the right data entered? Was all the data entered? By building validation into the templates and using 'double entry' techniques, the data validation step becomes much more manageable.

**Problems Addressed**

The rational was that templates and structural change should help address the following problems:
- permission management
- verification
- version control

This clear separation of the different roles of different staff members with respect to the spreadsheets designs out several major classes of common spreadsheet errors.

## 5   SUMMARY

The intention was that Hypernumbers should be a powerful new way for creating secure, error free and manageable applications with spreadsheet skills.

After a prolonged development process Hypernumbers is now being deployed in production for the first commercial clients. Details are unfortunately commercially confidential, but there are very promising results with large scale applications consisting of 10,000+ spreadsheet pages for use by tens of different users in different roles have been successfully created.

# 6 REFERERENCES

Raymond R Panko (2008) *What We Know About Spreadsheet Errors*
http://panko.shidler.hawaii.edu/My%20Publications/Whatknow.htm 2pm 21/4/2011

Heidi Roizen (2010) former CEO of T/Maker an early spreadsheet company for the Mac, personal communication

Stephen McCrory and Gordon Guthrie (2011) *Building Applications With Hypernumbers* Ecosse Press (forthcoming). Available for download from:
http://documentation.hypernumbers.com/contents/welcome/other-formats.html

Blaise Pascal (1656-1657), *Lettres Provinciales* no. 16.

Oasis Open Documents (2011) - see for instance the document OpenDocument v1.2 Committee Specification Public Review Draft 03, Part 2 Recalculated Formula (OpenFormula) Format.odt downloaded from http://www.oasis-open.org/committees/tc_home.php?wg_abbrev=office 5:30pm 21/4/2011

Simon Peyton Jones, Alan Blackwell, Margaret Burnett (2003) *A User-Centred Approach to Functions in Excel* http://research.microsoft.com/en-us/um/people/simonpj/papers/excel/excel.pdf 5:30pm 21/4/2011